\documentclass[12pt]{iopart}

 \usepackage{iopams} 
\usepackage[dvips]{graphicx}

\begin{document}

\title[Symmetry considerations on the magnetization 
process...]{Symmetry considerations on the magnetization process of 
the Heisenberg model on the pyrochlore lattice }

\author{
Karlo Penc$^1$, 
Nic Shannon$^2$, 
Yukitoshi Motome$^3$ and
Hiroyuki Shiba$^4$
}

\address{$^1$Research Institute for Solid State Physics and Optics, 
H-1525 Budapest, P.O.B. 49, Hungary. }
\address{$^2$H. H. Wills Physics Lab, Tyndall Av., Bristol BS8~1TL, 
UK}
\address{$^3$Department of Applied Physics, University of Tokyo, 
Bunkyo-ku, Tokyo 113-8656, Japan }
\address{$^4$The Institute of Pure and Applied Physics, 2-31-22 
Yushima, Bunkyo-ku, Tokyo 113-0034, Japan}

\begin{abstract}
We present a detailed symmetry analysis of the degeneracy lifting due 
to higher order spin exchanges in the pyrochlore lattice in applied 
magnetic field. Under the assumption of the 4 sublattice ordering, 
the criteria for a stable half-magnetization plateau are deduced. The 
higher order exchange terms may originate from spin--lattice 
coupling, or can describe quantum and thermal fluctuations.
\end{abstract}


\maketitle

\section{Introduction}

The very special feature of the antiferromagnetic nearest--neighbour Heisenberg model
on a pyrochlore lattice is that it does not order magnetically at 
{\it any} temperature~\cite{moessner,canals}.    
This is a consequence of the high (continuous) degeneracy of its ground state 
manifold, which for $O(3)$ spins is too great even for order--from--disorder 
effects to be effective.
None the less, the vast majority of magnetic materials with a 
pyrochlore lattice {\it do} order magnetically.
In these compounds, magnetic order is typically accompanied by a 
structural transition which lowers the crystal symmetry and so lifts the 
ground state degeneracy.
This ``order by distortion'' mechanism~\cite{tchernyshyov02,yamashita}
is a natural consequence of the magnetoelastic interactions present 
in almost all magnets.   
Examples may include YMn$_2$~\cite{canals,terao}, Y$_2$M$_2$O$_7$~\cite{keren}, but to date its most striking realization is in the high field 
properties of the chromium spinels CdCr$_2$O$_4$~\cite{ueda05} and 
HgCr$_2$O$_4$~\cite{ueda06}.

In these compounds a dramatic half--magnetization plateau is accompanied 
by a collossal magnetostriction and change of crystal symmetry.
The basic physics of this magnetization plateau can be understood by 
extending the simplest four--sublattice ``order by distortion''  model of 
the pyrochlore lattice~\cite{tchernyshyov02,yamashita} to finite magnetic field~\cite{penc04}. 
Here we develop a complete symmetry analysis of the most general 
four--sublattice order, and give additional details of the new phases 
which result when the assumptions made in~\cite{penc04} 
are relaxed to allow for more general spin--lattice interactions.

\section{The effective model}

We start by assuming a simplest microscopic form of magnetoelastic coupling,
in which the strength of exchange interactions between a pair of 
nearest neighbour classical spins $\langle i,j \rangle$ depends linearly on the 
change in length $\delta_{i,j}$ of the associated bond
\begin{equation}
 \mathcal{H} = \sum_{\langle i,j \rangle} 
  \left[
  J (1- \alpha \delta_{i,j}) {\bf S}_i\cdot {\bf S}_j
 + \frac{K}{2}  \delta_{i,j}^2 \right] - h \sum_{i} S_i^z \;.
 \label{eq:magnetoelastic}
\end{equation}
Here $\alpha$ is the spin-lattice coupling, and $K$ an elastic coupling constant.  
We choose  the z-axis to coincide with the direction of applied magnetic field.
As the experiments point to an ordered state, we further
assume that the simplest, four-sublattice order is stabilized, 
e.g. by longer range exchange interactions.  Since we are 
interested in static quantities, we can then safely reduce the  
problem to that of a single tetrahedron embedded in a pyrochlore lattice \footnote{In general, the changes in bond length $\delta_{i,j}$ are not independent 
variables, but rather implicit functions of the positions of all the magnetic ions.   
However, since in this special four--sublattice case the lattice is subject to a uniform distortion (an affine 
transformation), the lengths of the 6 bonds making up the tetrahedron can be treated 
as independent quantities.}. 

The symmetry group of this embedded tetrahedron is $\mathcal{T}_d$, which has 
24 elements and 5 irreps.  Bond variables such as $ \delta_{i,j}$ and 
${\bf S}_i \cdot {\bf S}_j$ transform according to the ${\sf A_1}$, ${\sf E}$ and ${\sf T_2}$ irreps:
\begin{equation}
 \left(
 \begin{array}{c}
  \bLambda_{{\sf A_1}}  \\
  \bLambda_{{\sf E},1}  \\
  \bLambda_{{\sf E},2}  \\
  \bLambda_{{\sf T_2},1}  \\
  \bLambda_{{\sf T_2},2}  \\
  \bLambda_{{\sf T_2},3}  \\
 \end{array}
\right)=
\left(
\begin{array}{cccccc}
\frac{1}{{\sqrt{6}}} & \frac{1}{{\sqrt{6}}} & \frac{1}{{\sqrt{6}}} &
 \frac{1}{{\sqrt{6}}} & \frac{1}{{\sqrt{6}}} & \frac{1}{{\sqrt{6}}} \\
\frac{1}{\sqrt{3}} & \frac{-1}{2\sqrt{3}} & \frac{-1}{2\sqrt{3}} &
 \frac{-1}{2\sqrt{3}} & \frac{-1}{2 \sqrt{3}} & \frac{1}{\sqrt{3}} \\
0 & \frac{1}{2} & -\frac{1}{2}  & -\frac{1}{2}  & \frac{1}{2} & 0 \\
0 & 0 & -\frac{1}{\sqrt{2}} & \frac{1}{\sqrt{2}} & 0 & 0 \\
0 & -\frac{1}{\sqrt{2}} & 0 & 0 & \frac{1}{\sqrt{2}}  & 0 \\
-\frac{1}{\sqrt{2}} & 0 & 0 & 0 & 0 & \frac{1}{\sqrt{2}} \\
\end{array}
\right)
\left(\begin{array}{c} 
 {\bf S}_{1} \cdot {\bf S}_{2} \\
 {\bf S}_{1} \cdot {\bf S}_{3} \\
 {\bf S}_{1} \cdot {\bf S}_{4} \\ 
 {\bf S}_{2} \cdot {\bf S}_{3} \\
 {\bf S}_{2} \cdot {\bf S}_{4} \\
 {\bf S}_{3} \cdot {\bf S}_{4} \\
\end{array}
\right) \;.
\label{eq:irreps}
\end{equation}
In this approach we assume that the the spin--space and the 
real--space are decoupled, i.e. that the L--S coupling 
and crystalline anisotropies are negligible.
\begin{figure}[ht]
  \centering
  \includegraphics[width=5.5truecm]{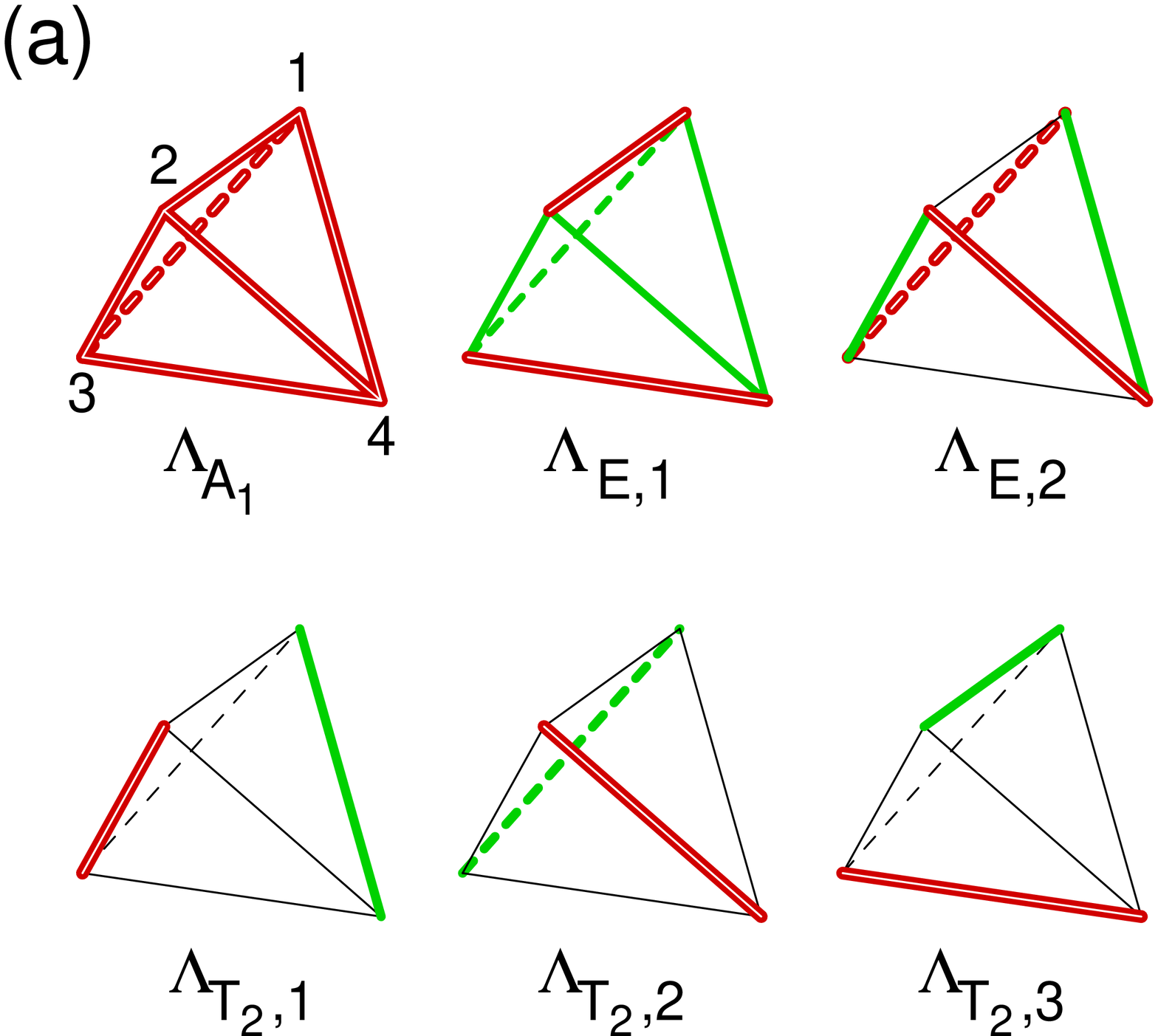}
  \hspace{1cm}
    \includegraphics[width=7.5truecm]{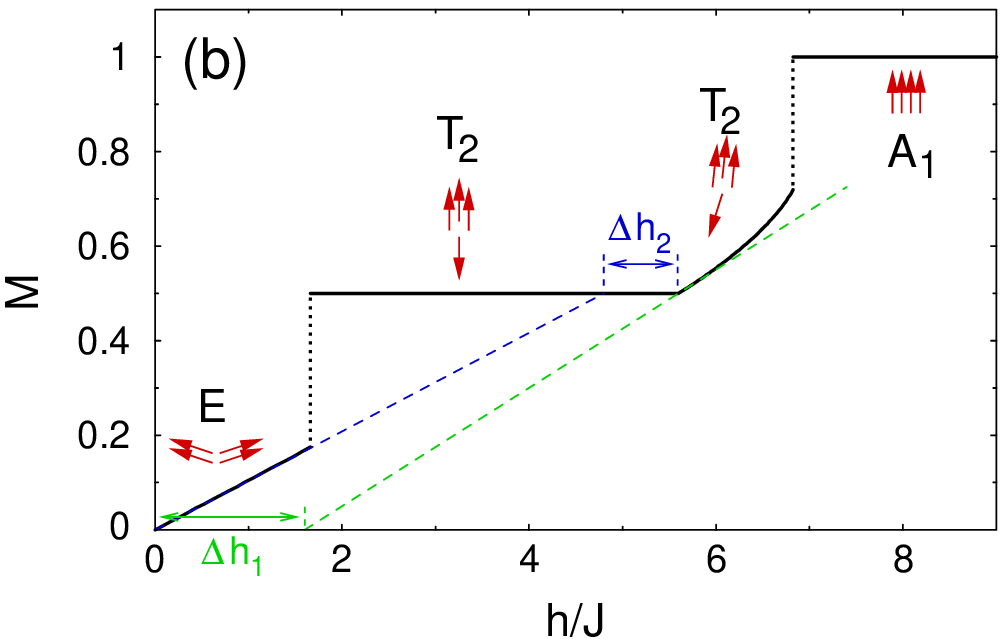}
  \caption{
  (a) Schematic representation of the different irreps of the tetrahedron. Hollow red lines have positive weight; green lines have negative weight; thin black lines have zero weight. 
(b) Magnetization curve for $b_{\sf A_1}=0.1$, $b_{\sf E}=0.1$ 
and $b_{\sf T_2}=0.2$. The spin configurations, together with the irrep they belong to are indicated.
  \label{fig:pd}}
\end{figure}

Under these assumptions, the Hamiltonian of the embedded tetrahedron 
is given by
\begin{eqnarray}
  \mathcal{H} &=& 2 \sqrt{6} J \Lambda_{\sf A_1} -4hM
   - 2  J 
   \left(
    \alpha_{\sf A_1} \Lambda_{\sf A_1} \rho_{\sf A_1} 
   +\alpha_{\sf E} \bLambda_{\sf E} \cdot  \brho_{\sf E}
   +\alpha_{\sf T_2}  \bLambda_{\sf T_2}\cdot  \brho_{\sf T_2}
   \right)\nonumber\\
   && +\left(
   K_{\sf A_1} \rho_{\sf A_1}^2 
   + K_{\sf E}  \brho_{\sf E}^2
   + K_{\sf T_2}  \brho_{\sf T_2}^2 \right) \;.
\end{eqnarray}
$M$ is the average magnetization per site.
The energy minima are found for $\rho_{\sf{R}}=(\alpha_{\sf{R}} 
J/K_{\sf{R}}) \Lambda_{\sf{R}}$, so the 
energy above becomes
\begin{equation}
  \mathcal{H} = 2 J \left( \sqrt{6} \Lambda_{\sf A_1}
   - b_{\sf A_1} \Lambda_{\sf A_1}^2
   - b_{\sf E}  \bLambda_{\sf E}^2
   - b_{\sf T_2}  \bLambda_{{\sf T_2}}^2 \right) -4hM
   \label{eq:HAET}
\end{equation}
with $b_{\sf R}=J \alpha_{\sf{R}}^2/(2 K_{\sf{R}})$. 
Actually, for the simple model defined by 
\Eref{eq:magnetoelastic}, all the couplings turn out to be equal: 
$b_{\sf A_1}=b_{\sf E}=b_{\sf T_2}$, and this is the case we have considered in Ref.~\cite{penc04}. 
This implied that only biquadratic terms of the form 
$({\bf S}_i\cdot {\bf S}_j)^2$ were present in the Hamiltonian. Once 
we allow nonequal couplings, we get three and four--site terms in the 
effective Hamiltonian of the form 
\begin{eqnarray}
\mathcal{H} &=&
2 J \sum_{i,j} {\bf S}_i\cdot {\bf S}_j - h \sum_i S_i^z
-\frac{J}{3} (b_{\sf A_1} + 2 b_{\sf E} +3 b_{\sf T_2})
\sum_{i,j} ({\bf S}_i\cdot {\bf S}_j)^2  
\nonumber\\
&&-\frac{2J}{3} (b_{\sf A_1} - b_{\sf E}) 
\sum_{i,j,k} ({\bf S}_i\cdot {\bf S}_j)({\bf S}_j\cdot {\bf 
S}_k)\nonumber\\
&&
-\frac{J}{3} (b_{\sf A_1} + 2 b_{\sf E} -3 b_{\sf T_2})
\sum_{i,j,k,l} ({\bf S}_i\cdot {\bf S}_j)({\bf S}_k\cdot {\bf S}_l) 
\;,
\end{eqnarray}
where the summations are carried out such that every possible 
combination of spins in the tetrahedron appears exactly once, furthermore $i\neq j\neq k \neq l$.

This result is of a very general nature: \Eref{eq:HAET} is the 
most general Hamiltonian at fourth order in spin operators 
describing a 4 sublattice long--range ordered state
on the pyrochlore or fcc--lattice.   
As such, it is independent of the actual origin of the coupling, 
which may equally be due to quantum and thermal fluctuations, or
higher order exchange processes~\cite{nikuni}. 
It could have been derived on purely phenomenologically grounds, as the 
sum of the group invariants ---  
$\Lambda_{{\sf A_1}}= 8(M^2-1/4)/\sqrt{6}$ (and its powers) is itself invariant 
while we can construct the 
quadratic forms $\bLambda_{\sf E}^2=\Lambda_{{\sf E},1}^2 + 
\Lambda_{{\sf E},2}^2$ and $\bLambda_{{\sf T_2}}^2=\Lambda_{{\sf 
T_2},1}^2 + \Lambda_{{\sf T_2},2}^2 + \Lambda_{{\sf T_2},3}^2$
from the ${\sf E}$ and ${\sf T_2}$ irreps. 
There are six cubic invariants; three trivial ones $\Lambda_{{\sf A_1}}^3$, $\Lambda_{{\sf A_1}} 
\bLambda_{\sf E}^2$, and  $\Lambda_{{\sf A_1}} \bLambda_{\sf T_2}^2$, 
and three nontrivial ones 
$\Lambda_{{\sf E},1}^3 -3 \Lambda_{{\sf E},1} \Lambda_{{\sf E},2}^2$, 
$\Lambda_{{\sf T_2},1} \Lambda_{{\sf T_2},2} \Lambda_{{\sf T_2},3}$ 
and
\begin{equation}
 \left(-\frac{1}{2} \Lambda_{{\sf E},1} 
   - \frac{\sqrt{3}}{2} \Lambda_{{\sf E},2}\right) 
 \Lambda_{{\sf T_2},1}^2 
+ \left(- \frac{1}{2} \Lambda_{{\sf E},1} 
   + \frac{\sqrt{3}}{2} \Lambda_{{\sf E},2} \right) 
 \Lambda_{{\sf T_2},2}^2 
+ \Lambda_{{\sf E},1} \Lambda_{{\sf T_2},3}^2 \;.
\nonumber
\end{equation}
These can also, in principle, be added to the Hamiltonian, but 
resulting proliferation of parameters would make the problem intractable, 
and for that reason we neglect them here. 
All of the invarients considered above are constructed from 
two--spin dot products: at 6th order more complicated spin 
expression may appear.

\section{Stability analysis of the collinear states}
 
Both the plateau and the fully polarized states are realized with 
collinear spin configurations, with spins pointing along the magnetic 
field. At certain critical values of magnetic field, these configurations 
cease to be favorable, and the spins cant to optimize their energy. 
This can be a continuous transition, when the magnetization smoothly 
changes from the fractional value, or a first order transition, in 
which case the magnetization jumps. Here we introduce a simple 
effective Hamiltonian to treat this effect, first for the fully 
polarized state, and then for the plateau. 

\subsection{Instability of the fully polarized state}

Small fluctuations about the fully polarized state can be 
characterized by
\begin{equation}
    {\bf S}_j = \left( 
   \eta_j,
   \xi_j,
   \sqrt{1-\eta_j^2-\xi_j^2} 
 \right) \;,
 \end{equation}
where $\eta_j$ and $\xi_j$ are small. The $\eta_j$ and $\xi_j$ are 
site variables, which transform according to the ${\sf A_{1}}$ and 
${\sf T_{2}}$ irreps\footnote{The bond variables, which are 
``squares'' of site variables, can transform as ${\sf A_1}$, ${\sf 
E}$ and ${\sf T_2}$ (as the product ${\sf T_2}\otimes {\sf T_2} =  
{\sf A_1} \oplus {\sf E} \oplus {\sf T_1} \oplus {\sf T_2}$).
}. Furthermore, since the model is isotropic, the magnetization 
always points along the magnetic field,  
$\sum_j \eta_j =\sum_j \xi_j =0$. This means that the ${\sf A_{1}}$ 
irrep is zero, and we need to worry about the ${\sf T_{2}}$ irrep 
only. Apart from the space-group symmetry, the magnetic field reduces 
the $O(3)$ spin--symmetry to $U(1)$, the rotation of spins around the 
axis parallel to the magnetic field. Instead of the real $\eta_j$ and 
$\xi_j$ quantities it is useful to introduce the complex $\eta_j + i 
\xi_j$ quantity:
\begin{equation}   
\begin{array}{lcrcrcl} 
  \eta_1 + i \xi_1 = \mu_1 +\mu_2 +\mu_3 \;,&\quad&
  \eta_3 + i \xi_3 = -\mu_1 +\mu_2 -\mu_3 \;,\\
  \eta_2 + i \xi_2 = \mu_1 -\mu_2 -\mu_3 \;, &&
  \eta_4 + i \xi_4 = -\mu_1 -\mu_2 +\mu_3 \;,
     \end{array}
\end{equation}
where the complex $\bmu=(\mu_1,\mu_2,\mu_3)$ transforms according to 
the 3--dimensional ${\sf T_{2}}$ irrep.

\begin{figure}[ht]
  \centering
  \includegraphics[height=7truecm]{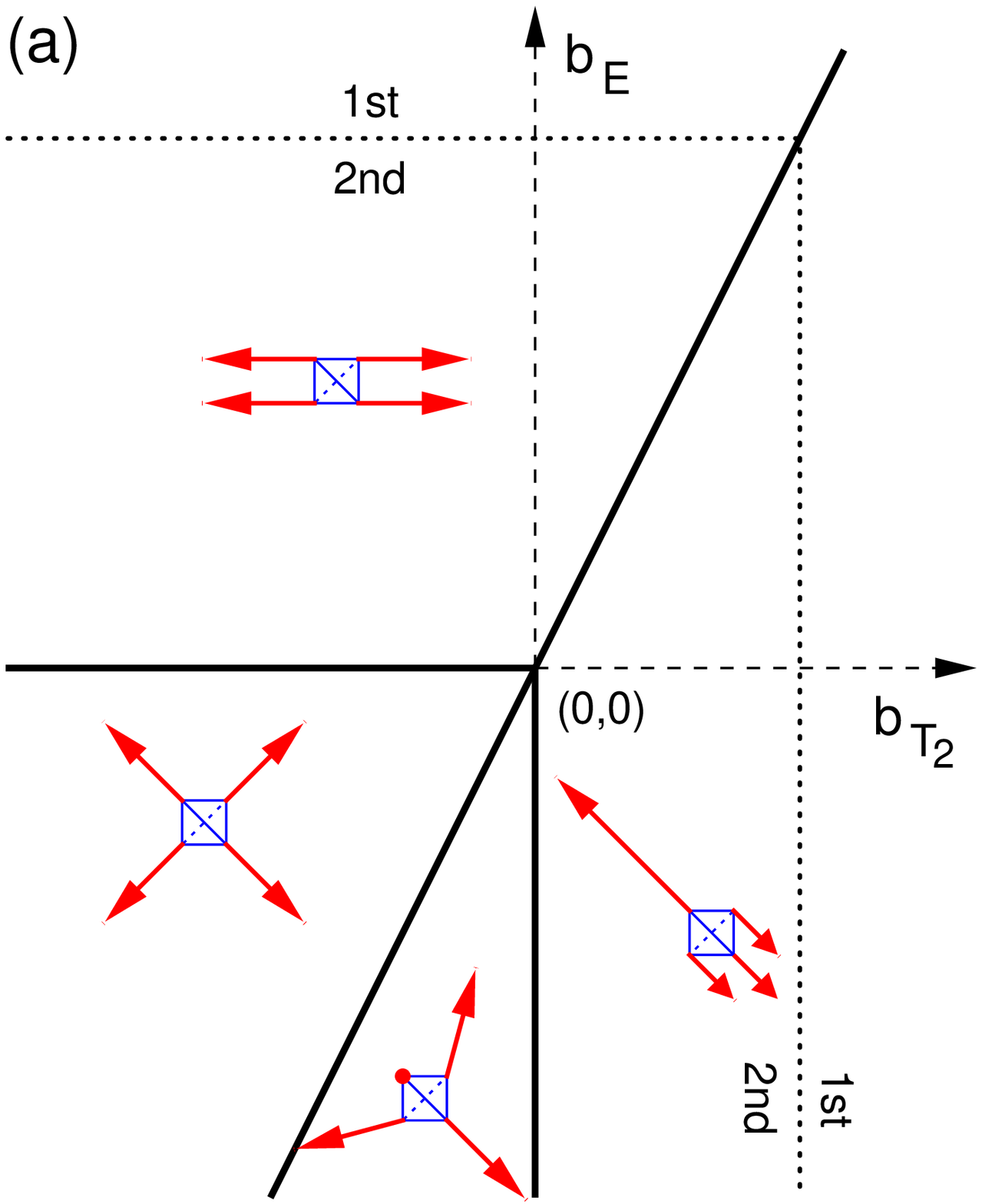}
  \hspace{0.3cm}
  \includegraphics[height=7truecm]{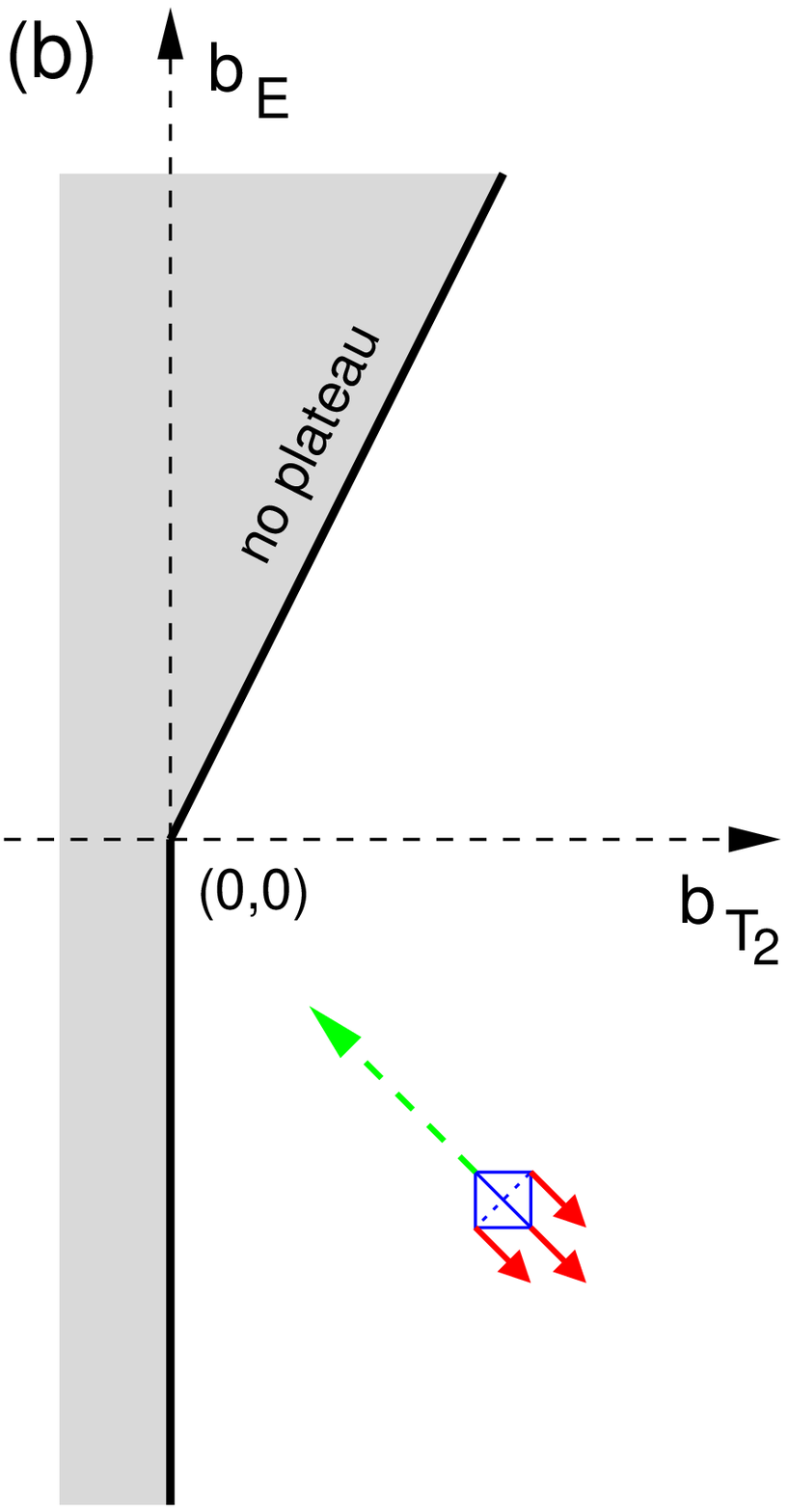}
  \hspace{0.3cm}
  \includegraphics[height=7truecm]{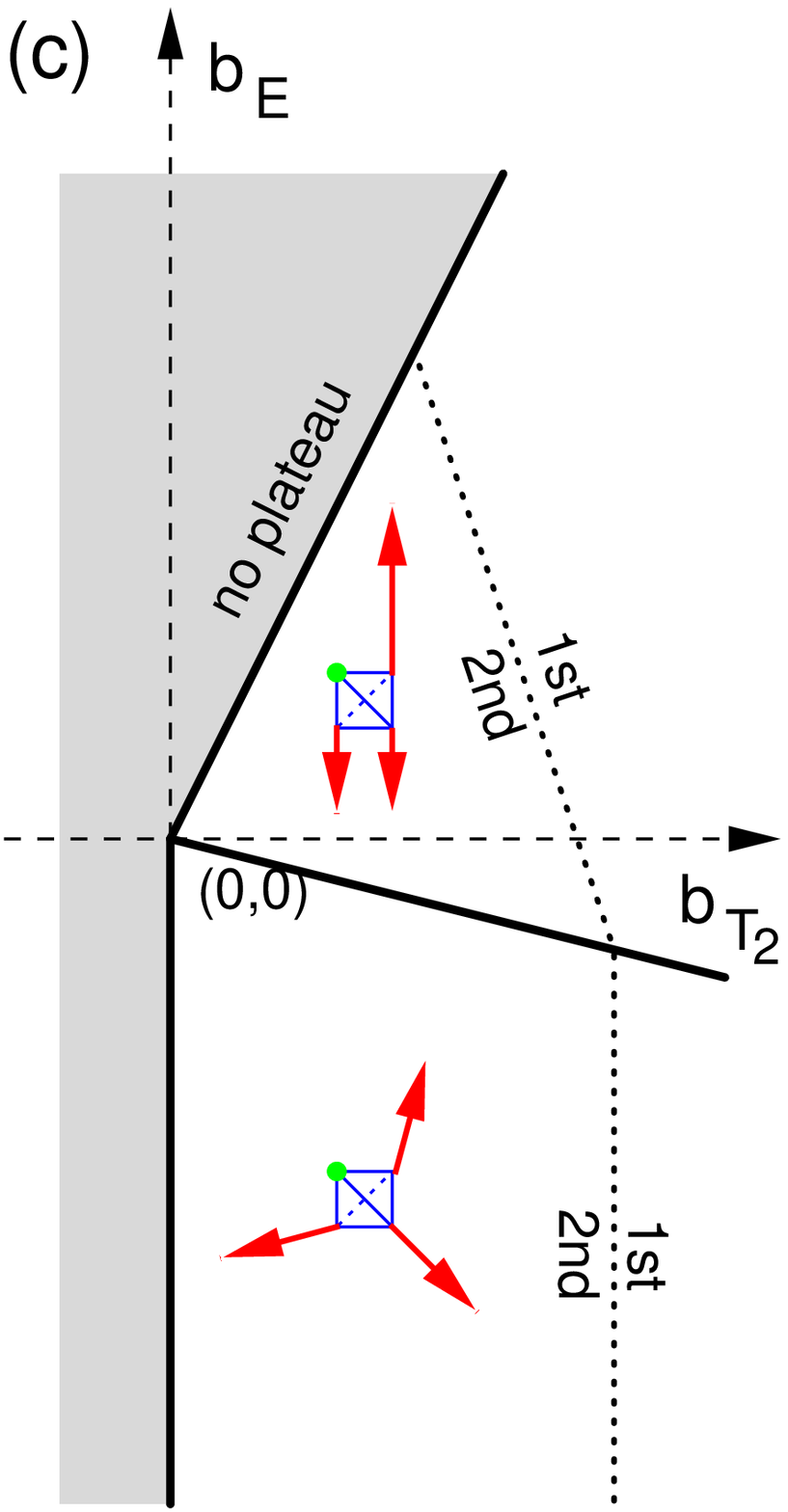}
  \caption{Instability of the (a) fully polarized state, and of the 
plateau at the (b) upper critical filed $h_u$ and (c) lower critical 
field $h_l$. The magnetic field is perpendicular to the plane, and 
the red arrows shows the spin canting $(\eta_j,\xi_j)$ of the $j$-th 
spin.
  Green dots and dashed arrows represent the down pointing spin in 
(b) and (c), while the shaded grey region where the plateau is not 
present.  \label{fig:FPD}}
\end{figure}

Assuming a second order transition, the fully polarized state becomes 
unstable at a critical field of 
\begin{eqnarray}
h_F=8 J(1 -2 b_{\sf A_1}) \;.
\end{eqnarray}
Expanding to fourth order in $\bmu$ at this critical field, the energy per spin of
the resulting canted state reads :
\begin{eqnarray}
  E &=& E_F + \frac{1}{2} (h-h_F) |\bmu|^2 
 + \frac{J}{3}\left(3-22 b_{\sf A_{1}} -8 b_{\sf E}\right)
  |\bmu|^4 
\nonumber\\ && 
 + 8 J (b_{\sf E}-b_{\sf T_{2}}) 
 \left( |\mu_1|^2 |\mu_2|^2 + |\mu_1|^2 |\mu_3|^2+ |\mu_2|^2 
|\mu_3|^2\right)
\nonumber\\ && 
 -4 J b_{\sf T_{2}}\left( 
 \mu_1^2 \overline{\mu}_2^2 +  \overline{\mu}_1^2 \mu_2^2 + 
 \mu_1^2 \overline{\mu}_3^2 +  \overline{\mu}_1^2 \mu_3^2 + 
 \mu_2^2 \overline{\mu}_3^2 +  \overline{\mu}_2^2 \mu_3^2 
 \right) \;,
\end{eqnarray}
  where $E_{F}=3J(1-b_{\sf A_{1}}) - h$ is the energy of a
  spin aligned with the magnetic field.
  Minimizing this energy as a 
function of the coupling parameters we obtain the results shown in 
\fref{fig:FPD}(a). In clockwise direction, we observe four 
phases: (i) The coplanar 2:2 canted phase of ${\sf E}$ symmetry with 
$\bmu\propto (1,0,0)$ (and permutations) when 
  $b_{\sf E}>0$ and $b_{\sf E}>2 b_{\sf T_2}$; (ii) The ${\sf T_2}$ 
symmetric coplanar 3:1 canted state with $|\mu_1|=|\mu_2|=|\mu_3|$ and equal arguments modulo $\pi$
for $b_{\sf T_2}>0$ and $b_{\sf E}<2 b_{\sf T_2}$; (iii) 
${\sf T_2}$ symmetric umbrella like configuration with $\bmu \propto 
(1,e^{\frac{2\pi i}{3}},e^{-\frac{2\pi i}{3}})$ when $b_{\sf T_2}<0$ and $b_{\sf 
E}<2 b_{\sf T_2}$
; (iv) An ${\sf E}$ symmetric 1:1:1:1 umbrella like configuration 
with $\bmu \propto (1,\pm i,0)$ when 
  $b_{\sf E}<0$ and $b_{\sf E}>2 b_{\sf T_2}$.

 The transition changes from a continuous into a first order one when 
the prefactor of the quartic term becomes negative. In the 2:2 phase, 
this happens for $8b_{\sf E} > 3-22 b_{\sf A_1}$, while in the 3:1 
state for $16b_{\sf T_2} > 3-22 b_{\sf A_1}$. Similarly first order 
transition takes place when $22 b_{\sf A_1} > 3-4 b_{\sf T_2}$ in 
case (iii) and $22 b_{\sf A_1} > 3-2 b_{\sf E}$ in case (iv).

\subsection{Instability of the plateau}

The analysis can be repeated for the stability of the plateau state: 
we assume that the spin at site 1 is down, ${\bf S_1} = \left( 
   \eta_1,
   \xi_1,
   -\sqrt{1-\eta_1^2-\xi_1^2} 
 \right)$
, while the others (sites 2, 3 and 4) are up.  Instabilities occur at 
\begin{equation}
  h_{l} = 4 J(1 - 4 b_{\sf T_{2}}) \quad {\rm and} \quad
  h_{u} = 4 J(1 + 2 b_{\sf T_{2}}) \label{eq:hu}\;,
\end{equation}
i.e. the plateau is stable for $h_{l}<h<h_u$, assuming a continuous 
phase transition, and its width is $24 b_{\sf T_{2}}$. Clearly first 
order transitions, if they occur, will shrink the magnitude of the plateau. 
Let us consider what happens at the upper edge of the plateau. 
The selection of the position of the down pointing spin breaks the symmetry in the 
$\bmu$ vector, and in our case, the soft mode is the $\bmu \propto 
(1,1,1)$ direction. It is convenient to introduce the $\mu^\| = 
(\mu_1+\mu_2+\mu_3)/\sqrt{3}$, and the $\mu^\perp_1 = 
(2\mu_1-\mu_2-\mu_3)/\sqrt{6}$ and $\mu^\perp_2 = 
(\mu_2-\mu_3)/\sqrt{2}$ orthogonal to the $\mu^\|$, the energy up to 
fourth order in $\mu$ (keeping the relevant quartic term only) reads
\begin{equation}
\fl
  E = E_P
  -\frac{1}{4} (h-h_u) |\mu^\||^2 
  + 12 J b_{\sf T_2} \left(|\mu^\perp_1|^2+|\mu^\perp_2|^2 \right)
  +\frac{J}{12}(3-4 b_{\sf A_1} + 2 b_{\sf T_2}) |\mu^\||^4 \;.
\end{equation}
 Here $E_P = -h/2 -3 J b_{\sf T_2}$ is the energy of the plateau. The 
instability is toward the 3:1 coplanar state, with the up spins 
having all equal canting angles. The continuous transition  becomes 
first order once  $3-4 b_{\sf A_1}+2b_{\sf T_2}<0$. 
This is the same 3:1 canted coplanar state found as the instability of the 
fully polarized solution, and there is no reason to suppose that the two 
states are not connected with each other - in other words, there is no
symmetry requirement for an additional phase transition when moving from the 
plateau to the fully polarized state\footnote{This need not be the case if 
we consider ordered states with more than four spins in the unit cell.}.

Next, let us consider what happens at the lower edge of the plateau: the energy is
\begin{eqnarray}
  E &=& E_P
  -\frac{1}{2} (h_l-h) \left(|\mu^\perp_1|^2+|\mu^\perp_2|^2 \right)
 + 6 b_{\sf T_2} |\mu^\||^2  
 \nonumber\\
&&  + \frac{2J}{3} \left( 4 b_{\sf E} + b_{\sf T_2} \right)  
  \left|\mu^\perp_1 \bar\mu^\perp_2 - \bar \mu^\perp_1 \mu^\perp_2 
\right|^2
  \nonumber\\ &&
  +\frac{J}{3} (3-4 b_{\sf A_1} - 8 b_{\sf E} - 24 b_{\sf T_2}) 
\left(|\mu^\perp_1|^2+|\mu^\perp_2|^2 \right)^2  
\end{eqnarray}
and, contrast to what happened at the upper critical field $h_u$, the softening 
occurs in the two modes perpendicular to $\mu^\|$.  
The $\left|\mu^\perp_1 \bar\mu^\perp_2 - \bar \mu^\perp_1 
\mu^\perp_2 \right|^2$ term selects the relative phase of the 
$\mu^\perp_1$ and $\mu^\perp_2$: $\mu^\perp_1=\pm i \mu^\perp_2$, 
and a ${\sf T_2}$ symmetric umbrella--like canting is realized for $4 
b_{\sf E} + b_{\sf T_2}<0$, while for $4 b_{\sf E} + b_{\sf 
T_2}>0$ the instability is against a degenerate coplanar canting 
(only the phase of the $\mu^\perp_1$ and $\mu^\perp_2$ is fixed; 
their relative amplitude $\mu^\perp_2/\mu^\perp_1$ is not).
In fact we need to go to 6th order to select the 2:1:1 state of 
mixed ${\sf E}$ and ${\sf T_2}$ symmetry previously found 
numerically~\cite{penc04}.   
This has three solutions:$(\mu_1,\mu_2)\propto(1,0)$, or 
$(-1/2,\pm\sqrt{3}/2)$. 
These transitions become first order  when 
$22 b_{\sf T_2} > 3-4b_{\sf A_1}$ for the umbrella state, and when $8 
b_{\sf E}+24 b_{\sf T_2} > 3-4b_{\sf A_1}$  for 2:1:1 state (see 
\fref{fig:FPD}(c)).

\section{Extrema of the invariants and magnetization curves}

Any given magnetization 
can be realized by a large number of (continuously degenerate) spin configurations. 
Within our model, the energy of these states
is uniquely determined by the second order invarients $\bLambda_{\sf R}^2$.
Therefore, by mapping out the extrema of these invariants, we exhaust all possible ground 
states selected by four spin interactions $b_{\sf R}$. 
Doing so enables us not only to confirm the results of the stability 
analysis above, but to make some further observations about the magnetization
process $M(h)$ and possible first order transitions.
   
\begin{figure}[ht]
  \centering
  \includegraphics[width=7truecm]{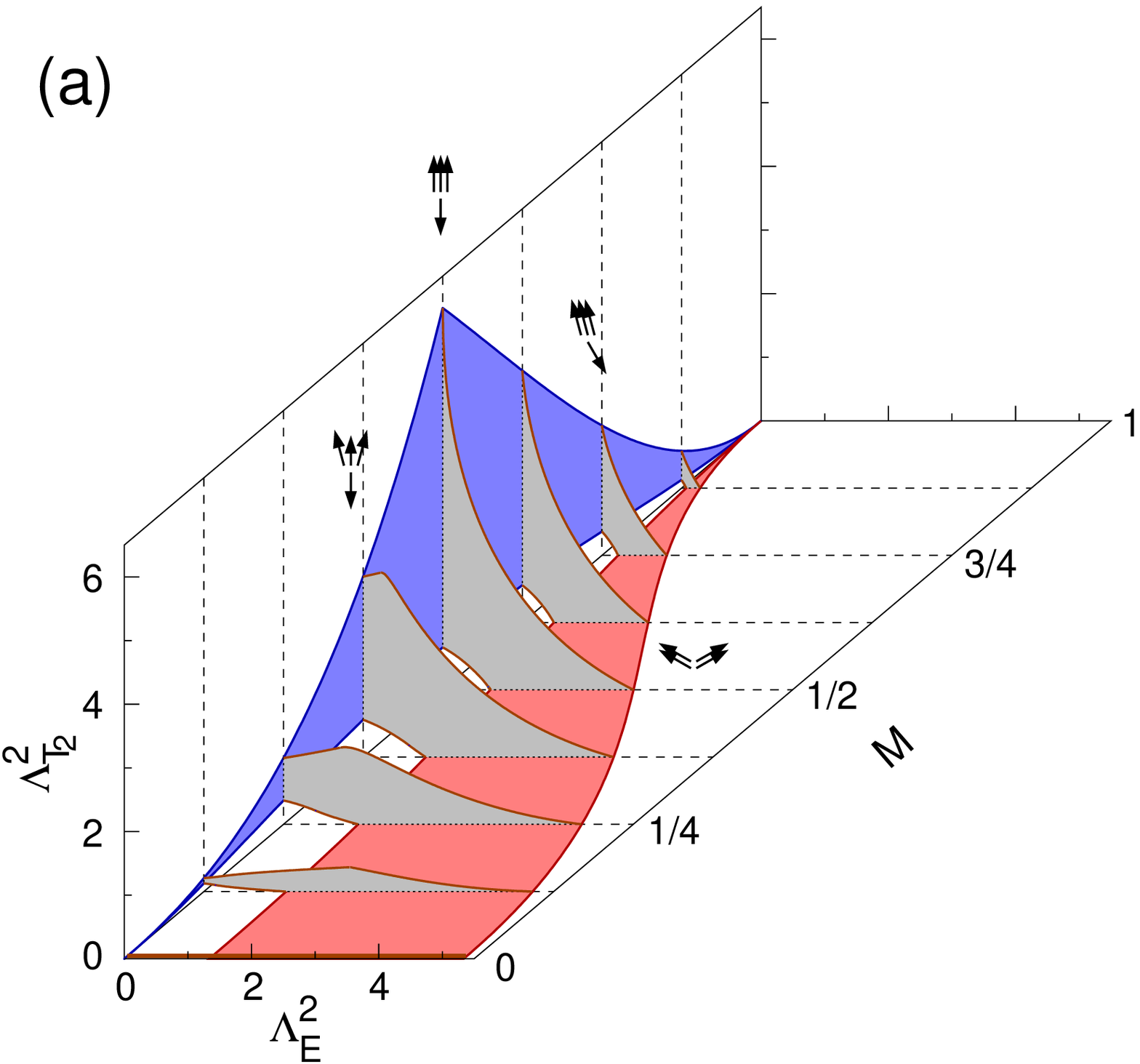}
      \includegraphics[width=6truecm]{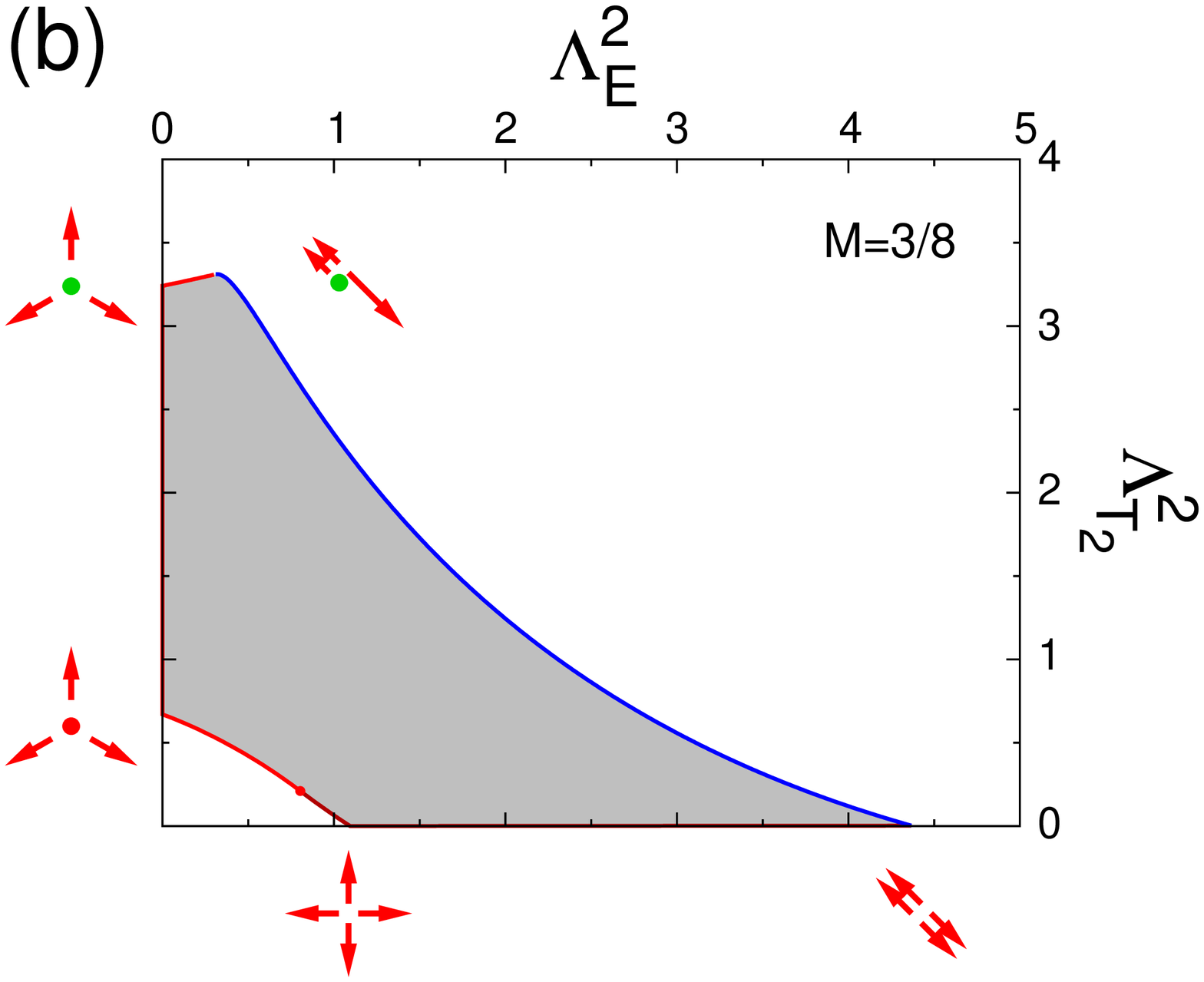}
  \caption{(a) The extremal values of $\bLambda_{\sf E}^2$ and 
$\bLambda_{\sf T_2}^2$ as a function of the magnetization $M$. (b) 
The cut at $M=3/8$. The degenerate manifold maps into the shaded 
region, and the border is outlined in blue for coplanar and red for 
non-coplanar states. The spin configurations are shown from the 
direction of the magnetic field, with red arrows canted in the 
direction of the field, and the green dots represent spins pointing 
opposite to the field.
  \label{fig:me2t2}}
\end{figure}

Let us first consider the limiting case $\bLambda_{\sf T_2}^2=0$.
Then, from \eref{eq:irreps}, the exchange energy on the opposing 
bonds of the tetrahedron is equal. This condition is satisfied by 
spins with equal $S^z$ components, and the $(S^x,S^y)=(\pm 1,0)$ and 
$(\pm \cos \phi, \pm \sin \phi)$. 
The maximum of value of $\bLambda_{\sf E}^2$ 
is obtained for a 2:2 canted state with $\phi=0$, while the minimum is 
obtained for a fully symmetric 1:1:1:1 canting with
$\phi=\pi/2$ (see \fref{fig:me2t2}), such that
\begin{equation}
\frac{4}{3} \left(1-M^2 \right)^2  \leq \bLambda_{\sf E}^2 \leq 
\frac{16}{3} \left(1-M^2 \right)^2 \;.
\end{equation}

Next, let us consider the case of $ \bLambda_{{\sf E}}=0$. The 
maximum value for $ \bLambda_{{\sf T_2}}^2=6$ is realized when all 
the $\Lambda_{{\sf T_2},i}$ take their maximal value $\Lambda_{{\sf 
T_2},i}=\sqrt{2}$ in the collinear uuud (three up--, one down--spin) 
configuration with $M=1/2$. 
For $M >1/2$,  $ \bLambda_{{\sf T_2}}^2$ 
is maximal for the coplanar 3:1 canted state.  
For $M<1/2$ the maximal state is a 
a three-dimensional inverted umbrella type configuration 
(${\bf S}_j=(\sin\vartheta \cos [j 2\pi/3], \sin \vartheta \sin [j 2 
\pi/3],\cos \vartheta)$ for $j=1,2,3$ and ${\bf S}_4=(0,0,-1)$,
while the minimal is obtained by inverting ${\bf S}_4$ to obtain a folded umbrella 
configuration, and
\begin{equation}
\frac{32}{3} M^2 (1-M)^2\leq  \bLambda_{{\sf T_2}}^2 \leq \left\{
 \begin{array}{cc}
 \frac{32}{3} \left(1-M^2 \right)^2 \;,
  & \mbox{if $1/2 \leq M \leq 1$\;;}\\
 \frac{32}{3} M^2 (1+M)^2 \;,
  & \mbox{if $0 \leq M \leq 1/2$\;.}
 \end{array}
 \right.
\end{equation}
However, a slightly better energy can be obtained by allowing finite values for 
$\bLambda_{{\sf E}}^2$, in which case the extremal point corresponds 
to coplanar 2:1:1 canted state with a relatively complex analytic form.

 The existence and stability of the plateau is related to the cusp of 
 $\bLambda_{\sf T_2}^2$ at $M=1/2$. As a singular point, it will 
 minimize the energy for a finite window of the magnetic field when 
 $2 b_{\sf T_2}>\max(b_{\sf E},0)$, also allowing for negative values 
 of $b_{\sf E}$.  This is in agreement with the results of the stability 
analysis presented above.  The extrema in the 
$(\bLambda_{\sf E}^2,\bLambda_{\sf T_2}^2)$ 
plane can clearly be related to the 
nature of instability of the plateau and the fully polarized state. 
 
 However, we can also see that for $b_{\sf E}>\max(2 b_{\sf T_2},0)$ 
the ${\sf T_2}$ order is not realized at all (here we allow for 
negative values of $b_{\sf T_2}$). The system remains in the coplanar
2:2 phase for all values of applied field up to saturation.
The half magnetization plateau is also absent when both 
$b_{\sf E}$ and $b_{\sf T_2}$ are negative: the ground state in this 
case is a noncoplanar spin configuration.  It is also possible to 
achieve a ${\sf T_2}$ type solution for small field, and an ${\sf E}$ 
type configuration for high field, with a first order transition between them.
  
In \fref{fig:pd}(b) we show the magnetization curve for a 
physically relevant set of parameters. To facilitate comparison with 
the experimentally measured curves, it is useful to extract the 
characteristic features of the magnetization curve.  
The energy per site of the 2:2 canted ${\sf E}$ symmetry state is: 
\begin{equation}
E_{2:2}=- h M 
- J (1-4 M^2) 
-\frac{J}{3} (1- 4 M^2)^2 b_{\sf A_1}
-\frac{8J}{3} (1 - M^2)^2 b_{\sf E}
\end{equation}
with zero field susceptibility
\begin{equation}
\left. \frac{\delta M}{\delta h} \right|_{h=0} 
 = \frac{3}{8J} \frac{1}{3 + 2 b_{\sf A_1} + 4 b_{\sf E} }\;.
\end{equation}
Similarly, the energy of the 3:1 canted state for $M\geq 1/2$ is : 
\begin{equation}
E_{3:1}=- h M
- J (1-4 M^2) 
-\frac{J}{3} (1- 4 M^2)^2 b_{\sf A_1}
-\frac{16J}{3} (1 - M^2)^2 b_{\sf T_2} \;,
\end{equation}
with differential susceptibility at the upper critical field:
\begin{equation}
\left. \frac{\delta M}{\delta h} \right|_{h=4 J +8 b_{\sf T_2}} = 
\frac{3}{8J} \frac{1}{3 - 4 b_{\sf A_1} + 2 b_{\sf T_2}} \;.
\end{equation}
If we continue the tangent of the magnetization curve at the upper 
critical field $h_u$ of the plateau, it will cut the $M=0$ axis at 
$h=16J(b_{\sf A_1}+b_{\sf T_2})/3$ ($\Delta h_1$ in 
\fref{fig:pd}(b)). Similarly, if we continue the linear 
magnetization at low field, it will intersect the plateau
at a distance
$\Delta h_2 = 8J (3 b_{\sf T_2} - b_{\sf A_1}- 2 b_{\sf E})/3$
from $h_u$. 
These values are easily extracted from the experimentally measured 
magnetization curves, and together with the upper critical field for 
the plateau, \eref{eq:hu}, allow to make a rough estimate of the 
strength of these parameters. Clearly, the cubic and higher order 
invariants will influence these results, and could also be used to fine 
tune the magnetization curve.
Furthermore, experimental evidence points toward more complicated, 8--
 \cite{chung05} or 16--sublattice \cite{huedapriv} ordered state, which is 
not described by this study. However, our preliminary results on the 
16 sublattice state seem to indicate that the shape of the 
magnetization curve may remain the same even in the case of 8 and 16 
sublattice ordering in certain (possibly relevant) cases, where the 
$b_{\sf R}$ parameters of the invariants of the $\mathcal{T}_d$ group are 
replaced by the appropriate coupling constant of the invariants of 
the larger point group. 

\section{Extreme quantum case}
\begin{figure}[ht]
  \centering
  \includegraphics[width=3.5truecm]{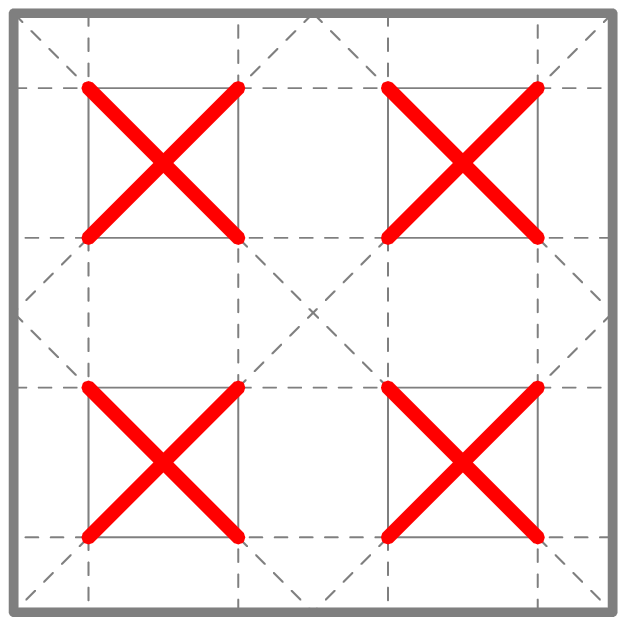}
  \includegraphics[width=3.5truecm]{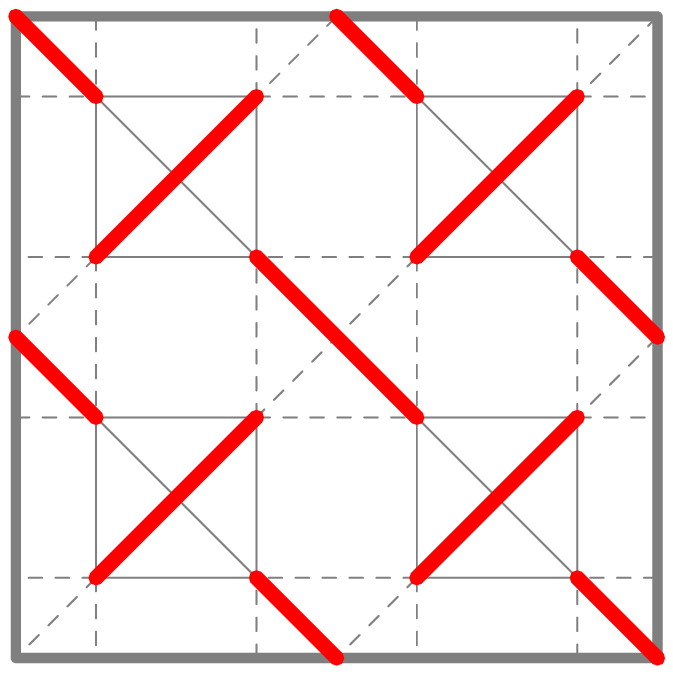}
  \caption{
\label{fig:dimer} Possible valence bond configurations which do not 
break translational symmetry.}
\end{figure}

In this paper we have concentrated on classical spins. What changes when we 
take into account the quantum nature of the spins? When the spins are 
ordered, we expect some corrections to the theory coming from quantum 
fluctuations (which are in fact effectively taken into account by the 
$b$'s \cite{nikuni}).  However, in spin-1/2 quantum limit, it may 
happen that spins chose to form singlets, rather than ordering 
magnetically  (similar physics for spin-1 has been studied in~\cite{yamashita}). 
Even though the unit cell of the ordered 
state remains the 4 site unit cell, we need to extend our theory to 
take into account the inversion symmetry $I$ broken by singlet bonds.
We therefore need 
to classify bond parameters according of the irreps ${\sf R'}$ of the 
cubic group $\mathcal{O}_h = \{1,I\}\otimes \mathcal{T}_d$.   Having 
done so, we can write an effective energy in terms of the $\bLambda_{\sf 
R'}^2$.  For zero magnetization, the possible valence bond patterns 
are shown in \fref{fig:dimer}: when $b_{\sf A_{2u}}+2 b_{\sf E_u} 
> 3 b_{\sf T_{1u}}$ \footnote{Using the standard notation for the irreps of 
the $\mathcal{O}_h$. The irreps ${\sf R}$ of the $\mathcal{T}_d$ become ${\sf R'}={\sf Rg}$ of the $\mathcal{O}_h$.},
 the valence bonds occupy alternate tetrahedra, otherwise they form a 
single valence bond in tetrahedron, with exchange energy 
distributed evenly among the tetrahedra.

\section{Conclusions}

In this paper, we have presented a complete symmetry analysis of all possible
four--sublattice phases of a classical pyrochlore antiferromagnet in applied
magnetic field, and determined its magnetic phase diagram for the most general
four--spin interactions.   This theory captures the essential physics
of spin--lattice coupling in spinel and pyrochlore materials, and
provides considerable insight into the magnetization plateaux observed
in Cr spinels.   In a separate publication, we will consider the new
complexities which arise when this symmetry analysis is generalized to
the full sixteen--site {\it cubic} unit cell of the pyrochlore lattice~\cite{penc-unpub}.

\section*{Acknowledgments}
We are pleased to acknowledge helpful discussions with M. Hagiwara, H. A. Katori, H. Takagi and H. Ueda.
This work was supported by the Hungarian OTKA Grants Nos. T049607 and K62280,
and EPSRC grant EP/C539974/1 and Grant No. 16GS50219 from the Ministry of
Education, Japan, as well as the HAS-JSPS bilateral project.
K.P. and N.S. are also grateful for the hospitality of MPI--PKS Dresden,
where a large part of this work was completed.


\section*{Bibliography}

\begin{thebibliography}{10}
    
\bibitem{moessner} Moessner R and Chalker JT 1998 {\it Phys. Rev.} B {\bf 58} 12049; Moessner R and Chalker JT
1998 {\it Phys. Rev. Lett.} {\bf 80} 2929

\bibitem{canals} Canals B and Lacroix C 2000 {\it Phys. Rev.} B {\bf 61} 1149; Canals B and Lacroix C 1998 {\it Phys. Rev. Lett.} {\bf 80} 2933

\bibitem{tchernyshyov02} Tchernyshyov O, Moessner R and Sondhi SL 
 2002 {\it Phys. Rev. Lett.} {\bf 88} 067203;
 Tchernyshyov O, Moessner R and Sondhi SL 2002 {\it Phys. Rev B} {\bf 66} 064403 

\bibitem{yamashita} Yamashita Y and Ueda K 2000 {\it Phys. Rev. Lett}. 
     {\bf 85} 4960
   
\bibitem{terao} Terao J 1996 {\it J. Phys. Soc. Jpn.} {\bf 65} 1413 

\bibitem{keren} Keren A and Gardner JS 2001 {\it Phys. Rev. Lett.} {\bf 87} 177201

\bibitem{ueda05} Ueda H, Katori HA, Mitamura H, Goto T and Takagi H 2005 {\it Phys. Rev. Lett.} {\bf 94} 047202
     
\bibitem{ueda06} Ueda H, Mitamura H, Goto T and Ueda Y 2006
       {\it Phys. Rev.} B~{\bf 73} 094415
       
\bibitem{penc04} Penc K, Shannon N and Shiba H 2004
		{\it Phys. Rev. Lett.} {\bf 93} 197203
		 
\bibitem{nikuni} Nikuni T and Shiba H 1993 {\it J. Phys. Soc. Jpn.}{\bf  62} 3268;
Shiba H, Nikuni T and Jacobs AE 2000 {\it J. Phys. Soc. Jpn.} {\bf 69} 1484 

\bibitem{chung05}   Chung JH, Matsuda M, Lee SH, Kakurai K, 
Ueda H, Sato TJ, Takagi H, Hong KP and Park S 2005 {\it Phys. Rev. 
Lett.} {\bf 95} 247204
	
\bibitem{huedapriv} Hiroaki Ueda, private communication

\bibitem{penc-unpub} Penc K {\it et al.}, in preparation


\end{thebibliography}
\end{document}